# Observation of Phasons in Magnetic Shape Memory Alloy Ni$_2$MnGa

S. M. Shapiro[1], P. Vorderwisch[2], K. Habicht[2], K. Hradil[3], H. Schneider[3]

[1]*Brookhaven National Laboratory, PO Box 5000, Upton, NY*

[2]*Hahn-Meitner Insititute, Berlin, Germany*

[3]*Georg-August-Universität, Göttingen, Germany*

An inelastic neutron scattering study of the lattice dynamics of the martensite phase of the ferromagnetic shape memory alloy, Ni$_2$MnGa, reveals the presence of well-defined phasons associated with the charge density wave (CDW) resulting from Fermi surface (FS) nesting. The velocity and the temperature dependence of the phason are measured as well as the anomalous [110]-TA$_2$ phonon.

It has been over 50 years since Peierls pointed out that in a one-dimensional metal where the electrons are coupled to the underlying lattice, the lattice is unstable at low temperature[1]. The ground state of this system has a charge density wave (CDW) where the electron density is modulated with a wavelength related to the Fermi wave vector, $k_F$ and characterized by a gap in the single-particle excitation spectra[2]. In order to maintain charge neutrality, the lattice will also be spatially modulated with wave-vector $2k_F$. CDW ground states were initially observed in a number of low dimensional solids because a large electronic susceptibility develops as a result of nesting of various flat pieces of the Fermi surface. This wave-vector is usually incommensurate with the lattice. The complex quantity that represents the lattice distortion accompanying the CDW may be taken as the order parameter of the phase transitions and has an amplitude and phase. The fluctuations in these quantities are called, respectively, amplitudons and phason. Because the energy of the system does not depend upon the phase of the order parameter it is considered the symmetry breaking Goldstone mode and is gapless. These excitations have been extensively studied over the last 30 years. The amplitude mode has been observed in the CDW state most easily by Raman scattering techniques[3], but there have been limited observations of well defined propagating phasons[4]. Incommensurate ground states have been observed in many nonmetallic systems that are also 3-dimensional and have been extensively studied [5]. For these systems the origin of the incommensurability is not related to the electronic configuration, but due to competition of various long range interatomic forces[5]. In this paper we report on the observation of well-defined phasons in the 3-D metallic compound, $Ni_2MnGa$, a material under intense study due to its ferromagnetic shape memory properties[6].

$Ni_2MnGa$ is a ferromagnetic shape memory alloy that exhibits the largest known magnetostrain effects (9.5%) and shows great promise for technical applications[7]. The stoichiometric compound is ferromagnetic below $T_C \sim 380$ K and exhibits two thermally induced phases at lower temperatures[8]. The high temperature phase is the cubic Heusler structure (a = 5.81Å) and a premartensitic transformation occurs at $T_{PM} \sim 260K$ which can be viewed as cubic with a 3-fold modulation along the [110] direction[9]. At a lower temperature, $T_M \sim 220K$, the crystal undergoes a martensitic transformation to a modulated tetragonal phase. The martensite phase is tetragonal with c/a=0.94 (a=5.90Å) with an incommensurate[10] modulation along the [110] direction consisting of shuffles of {110} planes along the perpendicular direction [$\bar{1}$10]. The modulation corresponds to nearly 5 interplanar distances but is actually incommensurate with a wavevector $q_{CDW}$=(0.43,043,0) as shown in the diffraction pattern of Fig. 1a. Early inelastic neutron scattering studies on the high temperature cubic phase showed a dramatic softening of the [110]-$TA_2$ phonon branch at the wavevector $\mathbf{q}_{PM}$=(0.33,0.33,0). This phonon softening is considered a precursor to the intermediate phase[11].

Recently there have been a number of first principles studies of the electronic structure and the lattice dynamics of $Ni_2MnGa$ whose aim was to explain the variety of phase transitions and modulated structures occurring in this system[12-15]. The most recent calculation[15] reproduced the experimental results of a phonon anomaly at $q_{PM}$ in the high temperature cubic phase and identified it as a giant Kohn anomaly resulting from a strong screening due to electron-phonon coupling and Fermi surface nesting. They also performed calculations in the martensite phase with volume preserving tetragonal distortion and varying c/a ratios. For the observed case of c/a = 0.94, they find strong

Fermi surface nesting is present at the wavevector $\mathbf{q}_{CDW}$= (0.43,043,0) and a phonon mode is unstable at the same wavevector. These features, an incommensurate modulation generated by Fermi surface nesting and a phonon anomaly, are the precise ingredients of charge density waves (CDW) that have been under intense study for over 30 years.

The single crystal of $Ni_2MnGa$ was cut from the same boule used in a previous experiment[10,11]. It was in the shape of half cylinder, 5 cm long and 1.5 cm diameter. The pre-martensitic transition temperature is 260K with little hysteresis. The martensitic transition temperature on cooling is $T_C$ = 212K and on heating $T_H$ = 228K. A great difficulty in studying the lattice dynamics of crystals in the martensite phase is the presence of many structural domains due to the lower symmetry of the martensite phase. It was shown, that by cooling in a modest magnetic field, a single domain is achieved due to magnetostrictive coupling. We recently reported on the TA modes in the martesite phase propagating along [110] and [101] directions[16]. The tetragonal c-axes of all the domains are forced to align parallel to the magnetic field. In our experiment we cooled the sample in a 1.5T vertical field resulting in a single domain sample with a (HH0) horizontal scattering plane.

The neutron scattering experiments were performed at the FRMII reactor at the Technical University-Munich and the Berlin Neutron Scattering Center at the Hahn Meitner Institute. At Munich, the PUMA spectrometer was used with a fixed final energy of 14.7 meV and collimations of 20'-24'-20'-30' between reactor and monochromator, monochramator and sample, sample and analyzer, and analyzer and detector respectively. Pyrolytic graphite (PG) was used as monochromator, analyzer and filter to remove higher order contamination of the beam. The resulting energy resolution for elastic scattering was $\delta E$=0.5 meV, full-width-half-maximum (FWHM). At Berlin, the FLEX instrument is situated on a cold source and we performed scans with a fixed final energy $E_f$ = 6.7 meV. This enabled higher resolution studies of the low energy excitations, $\delta E$=0.2 meV FWHM. Most of the measurements were performed in the (2,2,0) Brillouin zone using either constant-E or constant-Q scans.

Figure 1a shows an elastic scan along the transverse [$\bar{2}$20] direction perpendicular to the [220] direction. The intense, resolution limited peaks at q=±0.425 are the CDW peaks. The intensity of this peak is 1-2 orders of magnitude less than the Bragg peak inensity. Since Ni$_2$MnGa undergoes a first order transition the peak intensity is nearly temperature independent up to the transformation temperature and disappears rapidly as shown in Fig. 1b. Most of the measurements were performed in the martensite phase at 200 K, with limited measurements made at 100K and 50K to probe the temperature dependence of the excitation. Figure 2 shows a series of constant-Q scans (a-c) and constant-E scans (d-f) measured with high resolution on FLEX probing the [110]-TA$_2$ mode. The constant-Q scan at q=0.2 (Fig. 2a) shows a sharp peak E=2.5 meV. On increasing q to 0.3 (Fig. 2b), the peak position increases slightly and a shoulder develops on the low energy side. The entire spectrum could be fit with two Gaussians. On further increasing q (Fig. 2c), the lower energy peak increases in intensity and shifts to lower energies and the higher energy peak decreases in intensity. For measurements at larger q, the constant-Q scans also show two peaks. A compilation of the constant-Q scans measured with the moderate resolution on PUMA yields the intensity contours shown in Fig. 3. The lines are guides to the eye through the intensity maxima. One clearly sees the acoustic mode beginning at q=0 and exhibiting an anomalous shape for q>0.2 rlu. The elastic CDW peak is seen at $q_{CDW}$=0.425 and there is inelastic scattering emanating from this peak. This is shown more dramatically in the high resolution constant-E scans shown in Fig. 2d-f performed at E=1.0, 1.5 and 2.0 meV, respectively. There are peaks in q equidistant from $q_{CDW}$ that disperse to larger q-values as the energy increases. For q>$q_{CDW}$ this branch levels off and weakens in intensity as q approaches the zone boundary. For q<$q_{CDW}$ the branch disperses towards the TA$_2$ mode and may interact with it. This branch is the phason associated with the CDW predicted by first principles electronic calculations[15]

A plot of the low energy portion of the dispersion curves determined from constant-Q and constant-E scans measured with higher resolution on FLEX is shown in Fig. 4. Clearly seen are the TA$_2$ acoustic branch and the phason branch emanating from the CDW peak. This experimental result is strikingly similar to the schematic figure proposed by Overhauser for metals that have a CDW structure[17]. The velocity of sound determined from the initial slope of the TA$_2$ branch is $v_{TA2}$=1.6 x10$^5$ cm/sec which is twice as large as the sound velocity measured in the cubic phase[18]. The branches emanating from $q_{CDW}$ have a larger velocity: $v_{PH}$= 2.3 x 10$^5$ cm/sec. The TA$_2$ branch shows an increase in energy on decreasing the temperature but the phason mode is temperature independent. In addition to the observed phason there should also be an optic-like mode associated with fluctuations of the amplitude of the CDW. In $Ni_2MnGa$ it is tempting to associate the $TA_2$ mode in the vicinity of $q_{CDW}$ wavevector as the amplitude mode. However, the intensity contours of Fig. 3 suggest that there is some additional intensity in the valley between the phason branches at $q_{CDW}\pm q$ that could also be associated with a broadened amplitudon. The intensity contours also suggest that there is a strong interaction between the acoustic mode and the phason mode near $q\sim 0.3$. Both of these observations require more studies.

Although the prediction of phasons and their dispersion in CDW systems is over 30 years old, their observations are notoriously sparse [5,19]. To our knowledge there are only 3 or 4 systems where phonon anomalies associated with CDW's have been observed and even fewer where well-defined phasons have been observed[4]. The natural question arises as to why in $Ni_2MnGa$ the phasons appear so well defined and easily observed? One solution could be that the CDW is well ordered as evidenced by the resolution limited diffraction peak shown in Fig.1. In some CDW systems the peaks have shorter range correlations that limit the lifetime of the phasons and inhibit their observations. Another explanation could be that the intensity of the CDW peak is very weak in other systems compared to the Bragg peak and that would make the dynamics harder to observe. In $Ni_2MnGa$ the CDW peak is quite strong as shown in Fig. 1. A third, more fundamental consideration, is the exact nature of the CDW. Is it a uniform incommensurate modulation of the lattice or does it consist of locally commensurate waves separated by narrow domain walls, or discommensurations?[19,20] For the latter, the dynamics is related to soliton formation and the presence of well defined excitations is unclear. Since the phasons are well defined in $Ni_2MnGa$ the situation suggests that of a uniform incommensurate modulation of the lattice.

Acknowledgements:
The authors acknowledge the discussions with W.Ku, W. Yin, G. Xu, L Manosa, S. Kivelson, J. C. Lashley, M. Rice and L. Tanner. One of us (SMS) is grateful for the hospitality at FRMII and HMI. Work at Brookhaven is supported by the Office of Science, U. S. Department of Energy under Contract No. DE-AC02-98CH10886

Figure Captions:

Figure 1: (a) Elastic scan along the transverse direction about the (2,2,0) Brillouin zone center showing the CDW peaks at **q** =(-.425,.425,0)

Figure 2. Constant-Q scans measured along the $TA_2$ direction at (a) q=0.2, (b) q=0.3 and (c) q=0.4.. Constant-E scans at (d) E=1.0 meV, (e) E = 1.5 meV and (f) E=2.0 meV. Data taken at BENSC at Hahn-Meitner in Berlin

Figure 3. Intensity contours determined from a series of constant-Q scans measured on PUMA instrument at FRM-II reactor. Black areas were outside the scan limits. The lines are guides through the intensity maxima

Figure 4. Low energy phonon dispersions measured along the transverse [110] direction. Measurements were made on the FLEX instrument at Hahn Meitner Institute reactor.

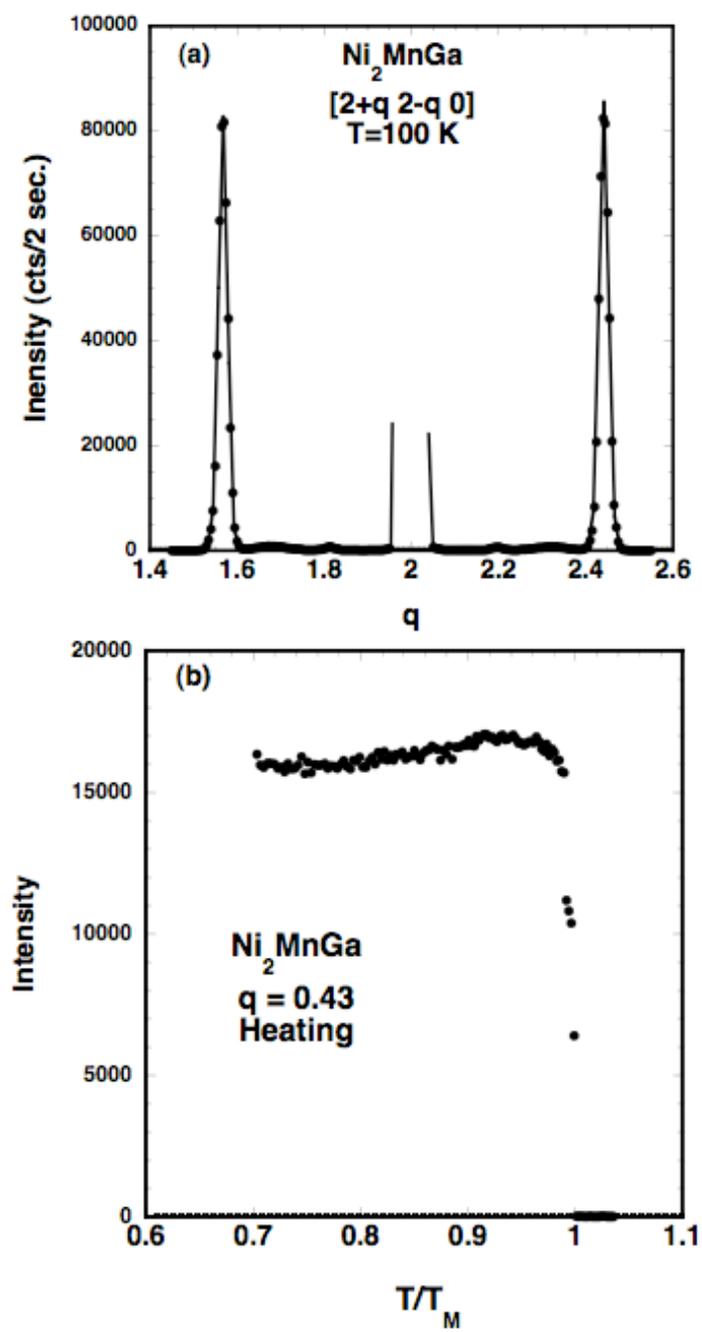

Figure 1.

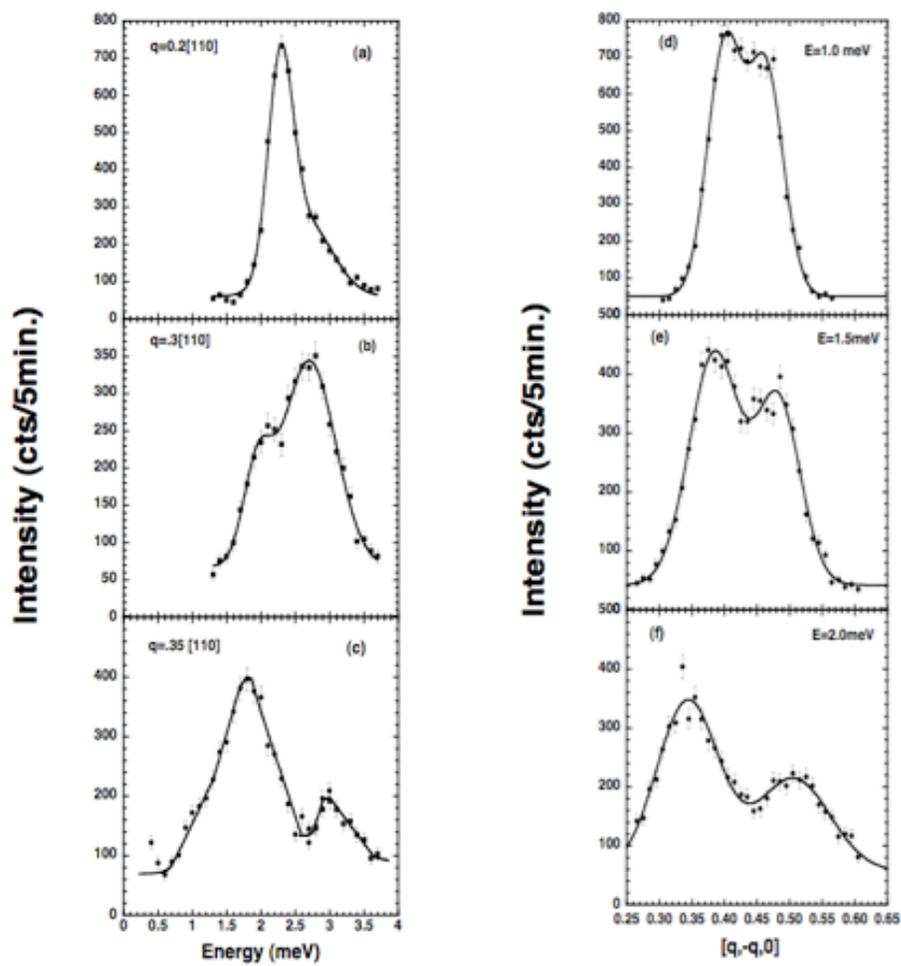

Figure 2.

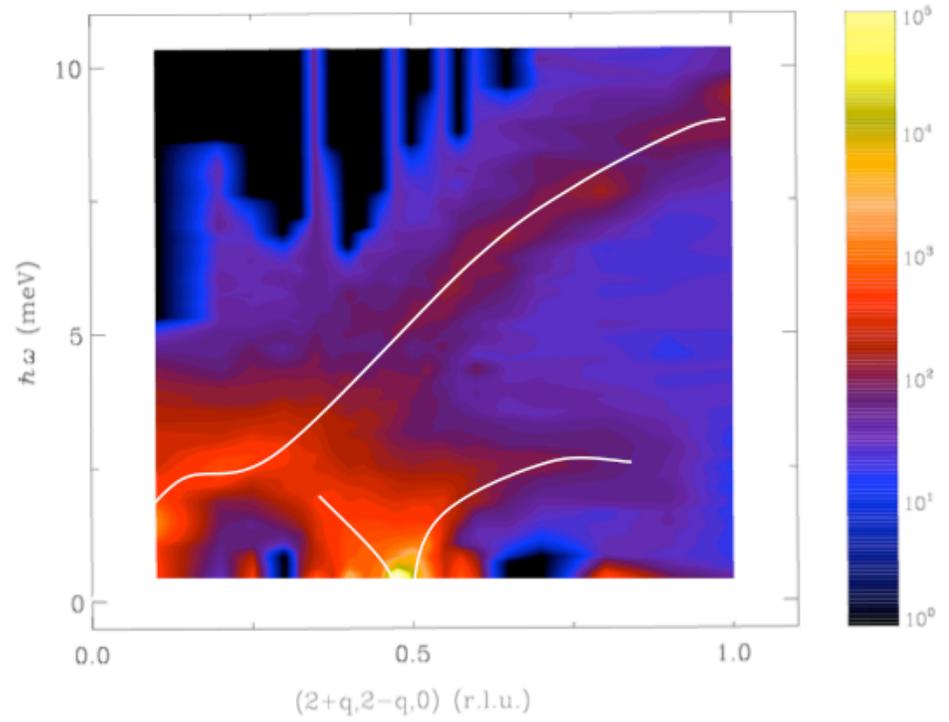

Figure 3

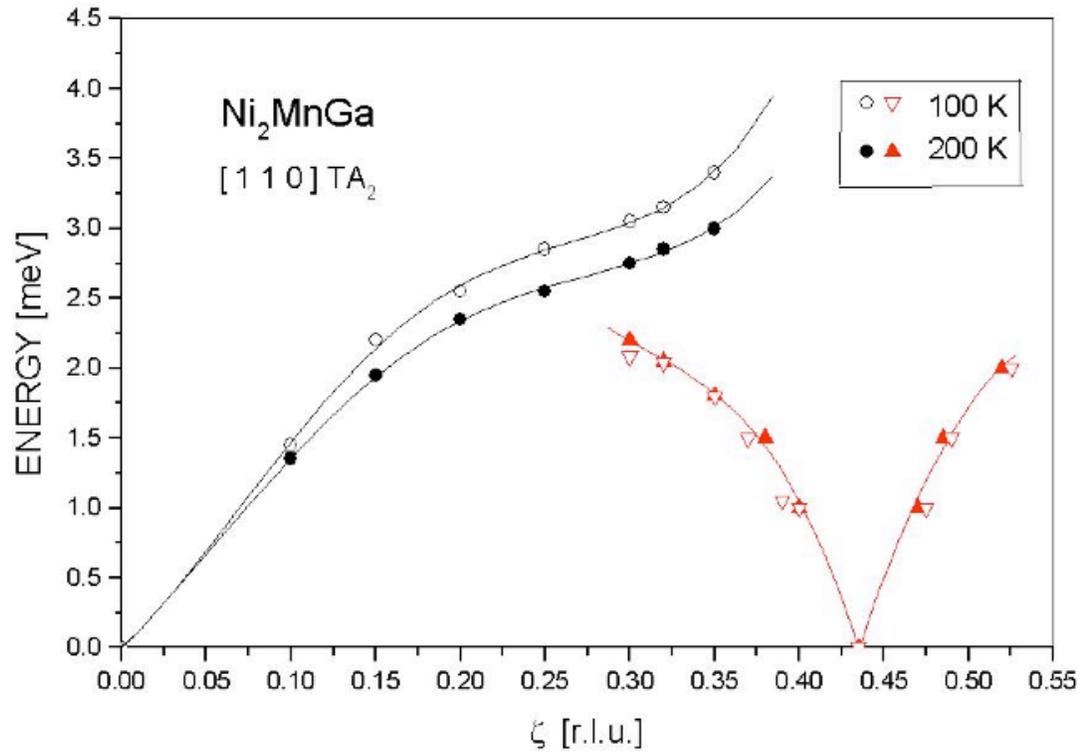

Figure 4